\documentstyle[nato]{crckapb}

\begin{opening}
\title{Covariant forms of Lax one-field operators: from Abelian to non-commutative}
\author{Sergey Leble}
\institute{\it Gda\`{n}sk University of Technology, ul.
Narutowicza 11/12, Gda\`{n}sk, Poland}
\end{opening}
\runningtitle{Covariant Lax one-field operators}
\begin{document}
\begin{abstract}
Polynomials in differentiation
  operators are considered. Joint covariance with respect to
   Darboux transformations of a
pair of such polynomials (Lax pair) as a function of one-field is
studied. Methodically, the transforms of the coefficients are
equalized to Frech\`{e}t differential (first term of the Taylor
series on prolonged space) to establish the operator forms.  In
the commutative (Abelian) case, as it was recently proved for the
KP-KdV Lax operators, it results in binary Bell (Faa de Bruno)
differential polynomials having natural bilinear (Hirota)
representation. Now next example of generalized Boussinesq
equation with variable coefficients is studied, the dressing chain
equations for the pair are derived. For a pair of generalized
Zakharov-Shabat problems a set of integrable (non-commutative)
potentials and hence nonlinear equations are constructed
altogether with explicit dressing formulas. Some non-Abelian
special functions are introduced.
\end{abstract}
\section {Introduction.}
Investigations of general Darboux transformation (DT) theory in
the case of differential operators
\begin{equation}\label{1}
L = \sum_{k=0}^n a_k\partial^k
\end{equation}
with non-commutative coefficients was launched by papers of
Matveev \cite{M}. The proof of a general covariance of the
equation
\begin{equation}\label{EQ}
 \psi_t = L\psi
\end{equation}
with respect to the classic DT (the shorthands $\psi' = \partial
\psi = \psi_x$ are used through the paper)
\begin{equation}\label{DT}
\psi[1]=\psi'-\sigma \psi,
\end{equation}
incorporates the auxiliary relation
\begin{equation}\label{Miu}
\begin{array}{c}
\sigma_t = \partial r  + [r,\sigma], \qquad
 r=\sum_0^Na_nB_n(\sigma),
\end{array}
\end{equation}
where $B_n$ are differential Bell (Faa de Bruno \cite{Faa})
polynomials \cite{LeZa}. The relation (\ref{Miu}) generalizes
so-called Miura map and became the identity when $\sigma =
\phi'\phi^{-1}$, $\phi$ is a solution of the equation (\ref{EQ}).

Such operators (\ref{1}) are used in the Lax representation
constructions for nonlinear  problems. It opens the way to produce
wide classes of solutions of the nonlinear problem.
 Examples of discrete, non-Abelian and non-local equations, integrable by DT
was considered in \cite{S}, \cite{LeSa}. Some of them were
reviewed and developed in the book \cite{MR93d:35136} and
intensely used nowadays \cite{RS}. The  approach was recently
generalized for a wide class of polynomials of automorphism on a
differential ring \cite{MR2002b:37115}.

A study of jointly covariant combinations introduces extra
problems of the appropriate choice of potentials on which the
polynomial coefficients depend \cite{MR93h:58142}. This problem
was recently discussed in \cite{Le}, where a method of the
conditions account was developed.
 Covariant combinations of (generalized) derivatives and potentials
may be hence classified for linear problems.
In two words, having the general statement about covariant form of
a linear polynomial differential operator that determines
transformation formulas for coefficients (Darboux theorem and its
Matveev's generalizations), the consistency between two such
formulas yields the special constraints. For example, the second
order scalar differential operator has the only place for a
potential and the covariance generate the classic Darboux
transformation for it.

In scalar case such one-potential constructions have been studied
in \cite{LLS} and developed for higher KdV and KP equations
\cite{LLoS}. It was found that the result is conveniently written
via such combinations of differentiation operator and exponential
functions of the potential as  Binary Bell Polynomials (BBP)
\cite{MR96m:58110}. The principle  is reproduced and developed in
the Sec. 2.1 of this paper to give more explanations.

The whole construction in general (non-Abelian) case is more
complicated, but much more rich and promising. The theory could
contain two ingredients.

i)The first one would be non-Abelian Hirota construction in the
terms of the mentioned binary Bell polynomials. On the level of
general formulation some obstacles appears, e.g. an extension of
addition formulas \cite{MR96m:58110}) to the non-Abelian case .

ii) The second way relates to some generalized polynomials that
could be produced as covariant combinations of  operators with a
faith that observations from Abelian theory could be generalized.
Namely the case we would discuss in this paper.

Even the minimal (first order in the $\partial$-operator) examples
of the ZS problems with operator coefficients  contains many
interesting integrable models. It is seen already from the point
of view of symmetry classification \cite{MR1776509}. So, the link
to DT covariance approach allows to hope for a realization of the
main purpose - construction of covariant functions, their
classification and use in the soliton equations theory.

We would begin from the example, using notations from quantum
mechanics to emphasize the non-Abelian nature of the
consideration.
 The operators $\rho$ and $H$ could play the roles of density
matrices and Hamiltonians, respectively, but one also can think of
them as just some operators without any particular quantum
mechanical connotations. The approach establishes the covariance
with respect to DT of rather general Lax system for the equation
 $$
 -i\rho_t=[H,h(\rho)],
 $$
where $h(\rho)$ - analytic  function, in some sense -"Abelian",
i.e. the function to be defined by Taylor series \cite{LCUK}.
  More exactly it is shown that the following statement takes place

  \medskip\noindent

{\bf Theorem.} {\it Assume $\langle\chi|$  and $\langle\psi|$ are
solutions of the following (direct) equations
$$
z_\nu\langle\chi|
 =
\langle\chi|(\rho -\nu H),\label{2-a}\\
$$
$$
-i\langle\chi_t| =
\frac{1}{\nu}\langle\chi| h(\rho) ,\label{2-b}\\
$$
and  $|\varphi\rangle$  stands for the conjugate pair.  Here
$\rho$, $H$ are operators left-acting on a ``bra" vectors
$\langle\psi|$ associated with an element of a Hilbert space. The
transforms   $\langle\psi_1|$, $\rho_1$, $h(\rho)_1$ are defined
by
\begin{equation}
 \langle\psi_1|
 =
\langle\psi|\Big(1+\frac{\nu-\mu}{\mu-\lambda}P\Big),\label{7.5.DT}
\end{equation}
\begin{equation}
\rho_1
 =
T \rho T^{-1},\label{7.5.DTU}\\
h_1(\rho) = T h(\rho)T^{-1}, \qquad T = \Big(
1+\frac{\mu-\nu}{\nu}P\Big).
\end{equation}
where
  $P =  |\varphi \rangle \langle \chi|\varphi \rangle \langle\chi|$.
Then the pairs are covariant:
$$
z_\lambda\langle\psi_1| = \langle\psi_1|(\rho_1 -\lambda H),
\qquad -i\langle\dot\psi_1| =
\frac{1}{\lambda}\langle\psi_1|h_1(\rho).
$$
complex numbers $\lambda$, $z_\lambda$ are independent of $t$
\cite{LCUK}.}

 The cases
$f(\rho)=i\rho^3$ and $f(\rho)=i\rho^{-1}$ were considered in
\cite{CKLN}, see applications in \cite{Minic}. A step to further
generalizations for essentially non-Abelian functions, e.g.  $h(X)
= XA+AX, [A,X] \neq 0$, is studied in \cite{MR2000c:82052}. The
case is the development of the matrix representation of the Euler
top  model \cite{MR56:13272}. This example of the theory  is more
close to the spirit of the sec.
 3.4, more achievements are demonstrated in \cite{CU},
 where abundant set of integrable equations is listed. The list is in a
partial correspondence with
 \cite{MR1776509},
 and give the usual for the DT technique link to solutions via the
iteration procedures or dressing chains.  One of the main results,
we present in the sec. 3.3, is how the "true" non-Abelian
functions appear in the context of the covariance conditions
application.

\section{One-field Lax pair for Abelian case.}
\subsection{ Covariance equations}
 First we would reproduce the "Abelian" scheme, generalizing the study of the example
  of the Boussinesq equation \cite{Le}. To start
 with the search we should fix the number of fields.
 Let us consider the third order operator (\ref{1}) with coefficients $b_k, k=0,1,2,3$,
 reserving $a_k$ for the second operator in a Lax pair.  Suppose, both
 operators depend on the only potential function $w$.
The problem we consider now may be formulated as follows: To find
restrictions on the coefficients $b_3(t)$,$ b_2(x,t)$, $b_1 =
b(w,t)$, $b_0 = G(w,t)$ compatible with DT transformations rules
of the potential function $w$ induced by DT for $b_i$.
 The classic DT for the third order operator coefficients (Matveev generalization
\cite{M}) yields
\begin{equation}\label{dt1}
b_2[1] = b_2 + b_3',
\end{equation}
\begin{equation}\label{dt2}
b_1[1] = b_1 + b_2' + 3b_3 \sigma',
\end{equation}
\begin{equation}\label{dt3}
b_0[1] = b_0 + b_1' +  \sigma  b_2' + 3b_3 (\sigma \sigma' +
 \sigma''),
\end{equation}
having in mind that the "elder" coefficient $b_3$  does not
transform. Note also, that $b_3'=0$ yields invariance of the
coefficient $b_2$.

 The general idea of DT form-invariance may be
realized considering the coefficients transforms to be consistent
with respect to the fixed transform of $w$. Generalizing the
analysis of the third order operator transformation \cite{Le}, one
arrives at the equations for the functions $b_2(x,t),b(w,t),G(w)$.
The covariance of the spectral equation
\begin{equation}\label{sp}
b_3 \psi_{xxx} + b_2(x,t)\psi_{xx} + b(w,t) \psi_{x} + G(w,t) \psi
= \lambda \psi
\end{equation}
may be considered separately, that leads to the link between $b_i$
only. We, however, study the problem of the (\ref{sp}) in the
context of Lax representation for some nonlinear equation, hence
the covariance of the second Lax equation is taken into account
from the very beginning. We name such principle as the "{\it
principle of joint covariance}" \cite{MR93h:58142}. The second
(evolution) equation of the case is :
\begin{equation}\label{ev}
\psi_t = a_2(t)\psi_{xx} + a_1(t)\psi_x +  w \psi,
\end{equation}
 with the operator in the r.h.s. having again the form of
 (\ref{1}).  We do not consider here a dependence of $a_i, b_i$  on
 $x$ for the sake of brevity, leaving this interesting question to the next paper.

   If one consider the $L$ and $A$ operators of the form (1), specified in equations
    (\ref{sp}) and (\ref{ev}).
as the Lax pair equations, the DT of $w$ implied by the covariance
of (\ref{ev}),
 should be compatible with DT formulas of
  both   coefficients of (\ref{sp}) depending on the only  variable
  $w$.
$$
a_2[1] = a_2 = a(x,t),
$$
$$
a_1[1] = a_1(x,t) + Da(x,t)
$$
\begin{equation}\label{DTa}
a_0[1] = w[1] = w +  a_{1}' + 2 a_2  \sigma' + \sigma a_2'
\end{equation}
Next important relations being in fact the identities in the
  DT transformation theory \cite{LeZa}, are the particular cases
  of the generalized Miura map (\ref{Miu}):
 \begin{equation}\label{M1}
 \sigma_t = [a_2( \sigma^2 + \sigma_x) +a_1\sigma + w]_x
\end{equation}
 for the problem (\ref{ev}) and, for the (\ref{sp})
\begin{equation}\label{M2}
\begin{array}{c}
b_3(\sigma^3 + 3\sigma_x\sigma + \sigma_{xx})+b_2( \sigma^2 +  \\
\sigma_x) + b(w,t)\sigma + G(w) = const;
\end{array}
\end{equation}
$\phi$ is a solution of both Lax equations.

 Suppose now that the coefficients
 of the operators are analytical functions of $w$ together with
 its derivatives (or integrals) with respect to $x$ (such functions are named functions
 on prolonged space \cite{O}).
 For the coefficient $b_0 = G(w,t)$ it means
 \begin{equation}\label{pro}
 G = G(\partial^{-1}w, w,
 w_x,...,\partial^{-1}w_t,w_t,w_{tx},...).
\end{equation}
The covariance condition is obtained for the Frech$\hat{e}$t
derivative (FD) of the function $G$ on the prolonged space, or the
first terms of multidimensional
 Taylor series for (\ref{pro}), read
\begin{equation}\label{FD}
\begin{array}{c}
  G(w +  a_{1}' + 2 a_2  \sigma' + \sigma a_2') = G(w) +  \\
 G_{w_x}(a_1'+ 2a_2\sigma'+ \sigma a_2')' + ... .
\end{array}
\end{equation}
We shall show only the terms of further importance.

Quite similar expansion arises for the coefficient $b_1 = b(w,t)$,
with which we would start
 in the analogy with the expressions (\ref{dt2}, \ref{FD}).
 Equalizing the DT and the expansion
  one obtains the condition
\begin{equation}\label{CC1}
 b_2' + 3b_3 \sigma' =  b_w(a_{1}' + 2 a_2  \sigma' + \sigma
a_2') + b_{w'}(a_{1}' + 2 a_2  \sigma' + \sigma a_2')' ... .
\end{equation}
This equation we name  the (first) {\it "joint covariance
equation"} that guarantee the consistency between transformations
of the coefficients of the Lax pair ((\ref{sp}),\ref{ev}).
 In the frame of our choice $a_2'=0$, the equation simplifies and
 linear independence of the derivatives $\sigma^{(n)}$ yields two
 constraints
\begin{equation}\label{CC11}
\begin{array}{c}
 3b_3  = 2b_w  a_2, \\
 b_2' = b_w a_{1}',
\end{array}
\end{equation}
or, solving the second and plugging into the first, results in
\begin{equation}\label{CC11'}
\begin{array}{c}
  b_w = 3b_3 /2a_2, \\
 b_2' =   3b_3 a_{1}'/2a_2.
\end{array}
\end{equation}
So, if one wants to save the form of
 the standard DT for the variable $w$
 (potential) the simple comparison of both transformation formulas gives for
 $b(w)$ the following connection (with arbitrary function $\alpha(t)$)
\begin{equation}\label{b}
  b(w,t)  = 3b_3w/2a_2 + \alpha(t).
\end{equation}

 Equalizing the expansion (\ref{FD}) with the
 transform of the $b_0=G(w,t)$ yields:
\begin{equation}\label{CC2}
\begin{array}{c}
  b_1' +  \sigma  b_2' + 3b_3 (\sigma^2/2  +
 \sigma')' =  \\
 G_{w_x}(a_1'+ 2a_2\sigma'+\sigma a_2')'
  + G_{\partial^{-1}w_t}[
  a_{1t}+2\partial^{-1}(a_2 \sigma_t')+ \partial^{-1}(\sigma a_2')_t]+...
\end{array}
\end{equation}
 This second {\it
 "joint covariance equation"} also simplifies when $a_2' = 0$:
\begin{equation}\label{CC2'}
\begin{array}{c}
  3b_3w'/2a_2 +  \sigma  b_2' + 3b_3 (\partial^{-1}\sigma_t-w
)'/2a_2 + 3b_3\sigma''/2 =
     G_{w_x}( a_1' + 2a_2\sigma')' + \\ G_{
\partial^{-1}w}[ a_1 + 2a_2 \sigma]+ G_{\partial^{-1}w_t}[ a_{1t} + 2a_2
\sigma_t] + ... ,
\end{array}
\end{equation}
 when (\ref{b}) is accounted. Note, that the "Miura" (\ref{M1}) is used in the
  l.h.s. and linearizes
 the FD with respect to $\sigma$. Therefore, the derivatives of the
 function $G$
\begin{equation}\label{G}
 \begin{array}{c}
     G_{w_x} = 3b_3/4a_2,\\
  G_{\partial^{-1}w_t} = 3b_3/4a_2^2,  \\
G_{\partial^{-1}w} = b_2'/2a_2,
 \end{array}
\end{equation}
are accompanied by the constraint
\begin{equation}\label{a1}
 a_{1t} + a_2a_1''+a_1a_1' = 0,
\end{equation}
which have got the form of the Burgers equation after
(\ref{CC11'}) account.  Finally the integration of the relation
(\ref{CC11'}) gives
\begin{equation}\label{b2}
b_2 = 3b_3a_1/2a_2 + \beta(t)
\end{equation}
and the "lower" coefficient of the third order operator is
expressed by
\begin{equation}\label{G}
G(w,t) = 3b_3w_x/2a_2 +  3b_3a_1'\partial^{-1}w /2(a_2)^2 +
3b_3\partial^{-1}w_t /2a_2^2.
\end{equation}
{\bf Statement 1} The expressions (\ref{ev}, \ref{sp}, \ref{b},
\ref{G}) define the covariant Lax pair when the constraints (
\ref{CC11'}, \ref{a1}) are valid.

{\bf Remark} We cut the Frech$\hat{e}$t differential formulas on
the level that is  necessary  for the minimal flows. The account
of higher terms leads to the whole hierarchy \cite{LLoS}.

\subsection{Compatibility condition.}

In the case $a_2' = 0$ by which we have restricted ourselves, the
Lax system (\ref{sp}, \ref{ev}) produces the following
compatibility conditions:
\begin{equation}\label{CoC}
\begin{array}{c}
2a_2b_3' = 3b_3a_2', \\
  b_{3t} = 2a_2b_2' - 3b_3a_1''\\
  b_{2t} = a_2b_2'' +2a_2b_1' +a_1b_2' - 3b_3a_1'' -  2b_2a_1'- 3b_3a_0'\\
  b_{1t} = a_2b_1'' + a_1b_1' - b_3a_1''' - b_2a_1'' -
   b_1a_1' - 3b_3a_0'' - 2b_2a_0' + 2a_2b_0'\\
    b_{0t} = a_1b_0'+ a_2b_0'' - b_1a_0'  - b_2a_0'' - b_3a_0'''
\end{array}
\end{equation}
In the particular case of $a_2  = 0$  we extract at once from the
first of the equalities (\ref{CoC}) the constraint $b_3' = 0$. The
direct corollary of (\ref{b2}) is $b_{3t} =0$.
  In the rest of
the equations the links (\ref{CoC},\ref{b2}) are taken into
account. Hence (\ref{a1}) in the combination with the expression
for $b_{2t}$ produce
\begin{equation}\label{beta}
  \beta_t = - 2\beta a_1'.
\end{equation}
The last two equations  (choice of constants $ b_3 =1, a_2 =
 -1 $)  become
\begin{equation}\label{bouss}
\begin{array}{c}
 \alpha w + \alpha_t + 3a_1''\partial^{-1}w/2 + (2\beta - 3a_1/2)w' + a_1''' + 3a_1a_1''/2   = 0 \\
    3 \partial^{-1}(w_t+a_1w)_t/4   =  (\alpha - 3w/2 )w' - w'''/4 + \\
   3 a_1w_t/4   + 3a_1a_1''\partial^{-1}w /4  +  3 a_1a_1'w/4 - 3a_1'w' /4 +\\
   (\beta + 3a_1/4 )w''.\
\end{array}
\end{equation}
In the simplest case of constant coefficients $(b_2' = a_1' = 0)$
one goes down to
\begin{equation}\label{B}
 \begin{array}{c}
  3b_3 (w_t+a_1w)_t/4a_2^2 =   \\
   -[(3b_3w/2a_2+\alpha)w' - b_3w'''/4 +
3b_3a_1w_t/4a_2^2 + (\beta - 3b_3a_1/4a_2)w'' ]'.   \
 \end{array}
\end{equation}
 This equation reduces to the standard Boussinesq equation when
 $(b_1 =a_1 = 0, b_3 =1, a_2 = -1)$  \cite{MR93d:35136}.

We would repeat that the results given in the Sec 2 are simplified
to show more clear the algorithm of the covariant Lax pair
derivation. More general study ($a_2' \neq 0$) will be published
elsewhere.

\subsection{Solutions. Dressing chains for the
Boussinesq equation}

The dressing formula for the zero seed potential (\ref{DT}) is
standard and  includes the only seed solution $\phi$,  of the Lax
equations with zero potential $w$.
\begin{equation}\label{sol1}
  w_s = a_{1}' + 2 a_2  \sigma' + \sigma a_2' = a_{1}'+ 2a_2\log_{xx}\phi(x,t)
\end{equation}

A next power tool to obtain solutions of nonlinear system is the
dressing chain equation: solitonic, finite-gap and other important
solutions were obtained for the KdV equations reducing such chain
\cite{W}. Going to the dressing chain, we use the scheme from
\cite{Le}.
 We would restrict ourselves further to
 the case of $a_2 = -1, a_0 = u, b_3 = 1, b_2 = 0 , b_1=b(u,t) = -3u/2 + \alpha,
 G = - 3u'/4+3\partial^{-1}u_t$ to fit the notations from \cite{Le}. The
 general construction is quite similar.

 The Miura equations (\ref{M1}, \ref{M2}) also simplifies:
\begin{equation}\label{M1s}
\sigma_t = -( \sigma^2 + \sigma_x)_x + u_{x}
\end{equation}
 for the problem (\ref{ev}) and
\begin{equation}\label{M2s}
  \sigma^3 + 3\sigma_x\sigma + \sigma_{xx} + b\sigma + G = const,
\end{equation}
where $b =  3u/2 + \alpha$, $G = -3\partial^{-1}u_t/4 + 3u_x/4$.

Namely the equations (\ref{M1s}, \ref{M2s}) together with the
n-fold iterated DT formula (\ref{7.5.DT})
\begin{equation}\label{CH1}
  u_{n+1} = u_n - 2  \sigma_n'
\end{equation}
form the basis to produce the DT dressing chain equations.

We express the iterated  potential $w_n$ from (\ref{M1s})
\begin{equation}\label{CH2}
 - \sigma_{nt} + ( \sigma_n^2 + \sigma_{n}')' = u_{n}'
\end{equation}
and substitute it into the differentiated relation (\ref{CH1}) to
get the first dressing chain equation
\begin{equation}\label{C}
\sigma_{n+1,t} - \sigma_{nt}  =  ( \sigma_{n+1}^2 +
\sigma_{n+1}')' - ( \sigma_n^2 - \sigma_{n}')'.
\end{equation}
Next chain equation is obtained when one plugs the potential from
(\ref{CH2}) to the iterated (\ref{M2s})
\begin{equation}\label{C2}
\sigma_n^3 + 3\sigma_n'\sigma_n + \sigma_n'' + (-3u_n/2 +
\alpha)\sigma_n + - 3u_n'/4+3\partial^{-1}u_{nt} = c_n.
\end{equation}

 \section{Non-Abelian case. Zakharov-Shabat (ZS) problem.}

\subsection{Joint covariance conditions for general ZS equations}
Let us change notations  for the first order (n=1) equation
(\ref{1}) with the coefficients from a non-Abelian differential
ring $A$ (for details of the mathematical objects definitions see
\cite{LeZa}) as follows
\begin{equation}\label{ZS1}
  \psi_t = (J+u\partial)\psi,
\end{equation}
where the operator $J \in A$ does not depend on $x,y,t$ and the
potential $a_0 \equiv u = u(x,y,t) \in A$ is a function of all
variables. The operator $\partial = \partial/\partial x$ may be
considered as a general differentiation as in \cite{LeZa}. The
transformed potential
 \begin{equation}\label{DT}
 \tilde{u} = u + [J,\sigma],
\end{equation}
where the $\sigma=\phi_x\phi^{-1}$, is defined by the same formula
as before, but the order of elements is important. The covariance
of the operator in (\ref{ZS1}) follows from general
transformations of the coefficients of a polynomial
\cite{MR93d:35136}. The coefficient $J$ does not transform.

Suppose the second operator of a Lax pair has the same form, but
with different entries and derivatives.
\begin{equation}\label{ZS2}
  \psi_y = (Y+w\partial)\psi,
\end{equation}
$Y \in A$ where the potential $w=F(u) \in A$ is a function of the
potential of the first (\ref{ZS1}) equation. The principle of
joint covariance \cite{MR93h:58142} hence reads:
  \begin{equation}\label{DT2}
 \tilde{w} = w+[Y,\sigma] = F(u+[J,\sigma]),
\end{equation}
with the direct corollary
 \begin{equation}\label{CP}
F(u)+[Y,\sigma] = F(u+[J,\sigma]).
\end{equation}
 So, the equation (\ref{CP}) defines the function
F(u), we shall name this equation as {\bf joint covariance
equation}. In the  case of Abelian algebra we used the Taylor
series (generalized by use of a Frechet derivative) to determine
the function. Now some more generalization is necessary. Let us
make some general remarks.

An operator-valued function  $F(u)$ of an operator $u$ in a Banach
space may be considered as a generalized Taylor series with
coefficients that are expressed in terms of Frech\`{e}t
derivatives.  The linear in u part of the series approximates (in
a sense of the space norm) the function
$$
F(u) = F(0) + F'(0)u + ... .
$$
The representation is not unique and the similar expression
$$
F(u)= F(0)+ u \hat{F}'(0)  + ...
$$
may be introduced (definitions are given in Appendix).  Both
expressions  however are not Hermitian, hence not suitable for the
majority of physical models. It means, that the class or such
operator functions is too restrictive. To explain what we have in
mind, let us consider examples.

\subsection{Important example}

From a point of view of the  physical modelling the following
Hermitian approximation
$$
 F(u) = F(0) + H^+u+uH + ...,\qquad u^+ = u,
$$
is preferable. Such models could be applied to quantum theories:
introduction of this approximation is similar to "phi in quadro"
(Landau-Ginzburg) model \cite{MR2000c:82052}. Let us study, in
which conditions the function
 \begin{equation}\label{LG}
 w = F(u)=Hu+uH,
\end{equation}
satisfy the joint covariance condition for the Lax pair
(\ref{ZS1}, \ref{ZS2})
 By direct calculation in (\ref{CP})
one arrives at the equality
 \begin{equation}\label{CP1}
 [Y,\sigma] = H[J,\sigma]+[J,\sigma]H.
\end{equation}
The obvious choice for arbitrary $\sigma$ is $Y=H^2, J=H$.

The compatibility conditions for the pair of equations (\ref{ZS1})
and (\ref{ZS2})  yields
\begin{equation}  \label{cc}
u_y-Hu_t-u_tH + [u,H]u+u[u,H] + H^2u_x+Hu_xH+H^2u_x = 0 \
\end{equation}

 If the potential does not
 depend on t, it is reduced to the next equation:
 \begin{equation}\label{GMan2}
   u_y  + [u^2,H]  + H^2u_x+Hu_xH+H^2u_x = 0,
\end{equation}
and x-independence yields the generalized Euler top equations
\begin{equation}\label{Euler}
   u_y  + [u^2,H] = 0,
\end{equation}
which  Lax pair  (\ref{ZS1},\ref{ZS2}) with $Y=J^2, J = H$ was
found by Manakov \cite{MR56:13272}.

\subsection{Covariant  combinations of symmetric polynomials}
The next natural example appears if one examine the link
(\ref{CP1}).
$$
P_2(H,u)=H^2u+HuH+uH^2
$$
The direct substitution in the covariance and compatibility
equations  leads to covariant constraint that turns to the
identity, if $Y=H^3,  J=H$.

It is easy to check more general connection $Y=J^n, J=H$
connection  that leads to the covariance of the function
$$
P_n(H,u)=\sum_{p=0}^nH^{{n-p}}uH^{p}.
$$
Such observation was exhibited in \cite{MR2000c:82052}. On the way
of a further generalization let us consider
\begin{equation}\label{comb}
f(H,u)=Hu+uH + S^2u+SuS+uS^2
\end{equation}
Plugging (\ref{comb}) as $F(u) = f(H,u) $ into (\ref{CP}),
representing $Y = AB + CDE$ yields
$$
A[B ,\sigma]+ [A ,\sigma]B+CD[E ,\sigma]+C[D ,\sigma]E+[C
,\sigma]DE=H[J ,\sigma]+[J ,\sigma]H +
$$
$$
S^2 [ J,\sigma]+ S[J
,\sigma]S+ [J ,\sigma]S^2.
$$
The last expression turns to identity if $A=B=J=H, C=\alpha H, D =
\alpha H, D = \alpha H, S = \beta H$  and $ [\alpha,
H]=0,[\beta,H]=0$ with the link $\alpha^3=\beta^2$.

   {\bf Statement} {\it Darboux covariance define a class of homogeneous
   polynomials $P_{n}(H,u)$, symmetric with respect to cyclic permutations. A linear
   combination of such polynomials $\sum_{n=1}^{N}\beta_nP_n(H,u)$ with the
    coefficients commuting with $u,H$ is also
   covariant, if the element $Y=\sum_{n=1}^{N}\alpha_n H^{n+1}$ and $\alpha_1=\beta_1=
   1, \alpha_n^{n+2}=\beta^{n+1}, n \neq 1 $.}

A proof could be made by induction that is based on homogeneity of
the $P_n$ and  linearity of the constraints with respect to u. The
functions $F_H(u) = \sum_0^{\infty} a_n P_{Hn}(u)$ satisfy the
constraints if the series converges.

\section{Conclusion}
  The main
result of this paper is the covariant equation (\ref{CP}). See,
also the example (\ref{CP1}). A class of potentials from \cite{CU}
contains polynomials $P_{n}(H,u)$ and give alternative expressions
 for it. The linear combinations, introduced here could better
 reproduce physical situation of interest. So, we used the compatibility
condition to find the form of integrable equation and reduction
 tracing the simplifications appearing for the subclasses of covariant potentials.
 While doing this we also check the invariance of the equation and  heredity of
the constraints.

The work is also supported by KBN grant 5P03B 040 20.

\appendix{Appendix. Right and left Frech$\hat{e}$t derivatives} The
classic notion of a derivative of an operator by other one is
defined in a Banach space $B$
 Two specific features  in the case of a
operator-function $F(u)\in\ B\,u\in B$  should be taken into
account: a norm choice when a limiting procedure is made and the
nonabelian character of expressions while the differential and
difference introduced.

{\bf Definition.} Let a Banach space $B$ have a structure of a
differential ring. Let $F$ be the operator from $B$ to $B'$
defined on the open set of $B$. The operator is named the
left-differentiable in $u _0 \in B$ if there exist a linear
restricted operator $L(u_0)$, acting also from B to B' with the
property
\begin{equation}\label{A}
 L(u_0+h)-L(u_0) = L(u_0)h +\alpha(u_0,h),\\
 ||h||\rightarrow 0,
\end{equation}
  where
$||\alpha(u_0,h)||/||h||\rightarrow 0.$ The operator
$L(u_0)=F'(u_0)$ is referred as the operator of the (strong) left
derivative of the function $F(u)$. The right derivative
$\hat{F}'(u_0)$ could be defined by the similar expression  and
conditions, if one changes $Lh \rightarrow h\hat{L}$ in the
equality (\ref{A}).

The addition of the half of the right and left differentials
\begin{equation}\label{FDs}
 ( F'(u_0) h+ h \hat{F}'(u_0))/2
\end{equation}
also approximates the difference   $L(u_0+h)-L(u_0)$  in the sense
of the FD definition.
\end{document}